\documentclass[journal, 9pt]{IEEEtran}

\usepackage{amscd,amsthm,amsmath,amssymb,amsfonts}
\usepackage[dvipsnames]{xcolor}
\usepackage{hyperref}

\hypersetup{
hidelinks,
colorlinks=true,
linkcolor=violet,
citecolor=NavyBlue,
}

\usepackage{cite}

\usepackage{graphics}
\usepackage{bm}
\usepackage{graphicx}
\usepackage{booktabs}
\usepackage{cryptocode}
\usepackage{url}
\usepackage{multirow}
\usepackage{makecell}

\newtheorem{theorem}{Theorem}
\newtheorem{fact}{Fact}
\newtheorem{remark}{Remark}

\title{Comments on ``Dynamic Consensus Committee-Based for Secure Data Sharing With Authorized Multi-Receiver Searchable Encryption''}

\author{Zi-Yuan Liu and Raylin Tso
\thanks{Zi-Yuan Liu and Raylin Tso are with the Department of Computer Science, National Chengchi University, Taipei 11605, Taiwan (e-mail: \{zyliu, raylin\}@cs.nccu.edu.tw)}
}

\begin{document}

\maketitle

\begin{abstract}
Recently, Yang \textit{et al.} introduced an efficient searchable encryption scheme
titled ``Dynamic Consensus Committee-Based for Secure Data Sharing With Authorized Multi-Receiver Searchable Encryption (DCC-SE),'' published in IEEE Transactions on Information Forensics and Security (DOI: 10.1109/TIFS.2023.3305183). According to the authors, DCC-SE meets various security requirements, especially the keyword trapdoor indistinguishability against chosen keyword attacks (KT-IND-CKA). In this letter, however, we reveal a significant vulnerability of DCC-SE: any users involved in the system can execute attacks against KT-IND-CKA security. This flaw potentially results in the unintended disclosure of sensitive keyword information related to the documents. We present a detailed cryptanalysis on DCC-SE. In addition, to address this vulnerability, we discuss the root cause and identify a flaw in the security proof of DCC-SE. Subsequently, we provide a solution that effectively addresses this concern without significantly increasing computational overhead.
\end{abstract}


\section{Introduction}
\IEEEPARstart{P}{ublic-key} searchable encryption, initially introduced by Boneh \textit{et al.}~\cite{DBLP:conf/eurocrypt/BonehCOP04}, has garnered significant attention due to its ability to search over ciphertexts. There have been various improvements in security and efficiency in this field. Among recent developments, Yang \textit{et al.}~\cite{DBLP:journals/tifs/YangTZH23} introduced an efficient authorized multi-receiver searchable encryption scheme (DCC-SE) based on the assistance of a consensus committee. As detailed in~\cite{DBLP:journals/tifs/YangTZH23}, DCC-SE offers high efficiency and provides several critical and flexible user management functionalities. It also satisfies security properties about the keyword information over ciphertexts and trapdoors. Moreover, the corresponding security proofs are also provided.

However, this letter highlights a significant flaw in DCC-SE: It actually does not satisfy the security called keyword trapdoor indistinguishability against chosen keyword attacks (KT-IND-CKA), contrary to the authors' claims. Specifically, in DCC-SE, sensitive keyword information within trapdoors may be compromised. The root cause is that any user, including malicious ones, who have registered in the system, receives the same secret token $\tau$ and, consequently, the same partial key $\eta$. Possessing the partial key $\eta$, any user that has obtained a trapdoor can adaptively generate ciphertexts for any random keyword and determine whether the ciphertext and the trapdoor correspond to the same keyword.
We provide a formal cryptanalysis of DCC-SE, demonstrating that how a malicious user can execute such attacks, following the security model of KT-IND-CKA. To resolve this issue, we suggest that the authors can limit the execution of the \textbf{Test} algorithm to a designated tester only. Therefore, any user cannot attempt to extract keyword information from trapdoors through testing.

\section{Review of DCC-SE}
\label{review}

Due to the page limitation, please refer to~\cite{DBLP:journals/tifs/YangTZH23} for comprehensive details of DCC-SE. In this letter, we only provide the Theorem~\ref{theorem} about  KT-IND-CKA security of DCC-SE, described in~\cite{DBLP:journals/tifs/YangTZH23}.

\begin{theorem}[KT-IND-CKA Security of DCC-SE (Theorem 2 in~\cite{DBLP:journals/tifs/YangTZH23})]
\label{theorem}
    If a PPT adversary $\mathcal{A}$ can break KT-IND-CKA security of DCC-SE, there is a PPT adversary $\mathcal{B}$ who can break the security of BLS blind signature and the DDH problem in the random oracle.
\end{theorem}
\section{Cryptanalysis of DCC-SE}
\label{cryptanalysis}

In this section, we present a cryptanalysis of DCC-SE. More specifically, we consider the scenario where a receiver $\mathcal{R}$ intends to search ciphertexts related to a data sender $\mathcal{S}$ and a keyword $w$, and therefore upload a trapdoor to the cloud via the public channel. We demonstrate that any (malicious) user that has registered in the system has the ability to obtain the keyword information associated with this trapdoor.\footnote{We here note that since $\mathcal{S}$ is the data sharer who shares data to $\mathcal{R}$ and is capable of generating valid ciphertexts $C_w$ to obtain the keyword information of $T$, we do not categorize $\mathcal{S}$ as the malicious user.}  Essentially, these malicious users are capable of compromising the KT-IND-CKA security of DCC-SE. Here, the ``registration'' of a user implies two conditions are met: \begin{enumerate}
    \item The user has received the secret token $\tau$ for the current epoch $\mathcal{J}$ from the certificate authority (followed by the \textbf{Setup} algorithm).
    
    \item The user has also obtained the partial key $\eta$ for the current epoch $\mathcal{J}$ from the committed nodes (followed by the \textbf{Partial Key Request} algorithm).
\end{enumerate} 

Let $\mathcal{J}$ be the current epoch. For simplicity, assume that the system comprises several users, including only one malicious user $\mathcal{A}$, with all having completed their registration. Now, since all users have registered in the system, they must receive the same secret token $\tau$ for the epoch $\mathcal{J}$ from the certificate authority. We can establish Fact~\ref{fact1}.

\begin{fact}
\label{fact1}
    All users, including the malicious user, possess identical information regarding the secret token $\tau$.
\end{fact}

Possessing the secret token $\tau$, all users (including the malicious user) can request the partial key $\eta$ from the committee nodes. As described in the \textbf{Partial Key Request} algorithm, the users will receive $\eta = H_2(\sigma_u)$, where $\sigma_u = s \cdot H_1(\tau)$ and $s$ is the secret key of the committee nodes. Consequently, this leads to the establishment of  Fact~\ref{fact2}.

\begin{fact}
\label{fact2}
All users, including the malicious user, possess identical information regarding the partial key $\eta$.
\end{fact}

\noindent \textbf{Our Attack.} Now, utilizing Fact~\ref{fact1} and Fact~\ref{fact2}, we demonstrate how the malicious user $\mathcal{A}$ attacks DCC-SE within the revised\footnote{In the original security model of KT-IND-CKA (defined by Yang \textit{et al.} in \cite{DBLP:journals/tifs/YangTZH23}), $\mathcal{B}$ does not provide $\tau$ and $\eta$ to $\mathcal{A}$. The model violates Fact \ref{fact1} and Fact \ref{fact2} we have discussed, causing the flaw in their security proof (See Section \ref{Discussion}).} KT-IND-CKA security model. Here, let $\mathcal{B}$ be the challenger who simulates the whole system for $\mathcal{A}$ as follows.

\noindent \underline{\textit{Setup.}} $\mathcal{B}$ provides the public key $\mathcal{X}$ of a data sender $\mathcal{S}$ and the public keys $\mathcal{Y}_{i=1,\cdots,n}$ of receivers $I := \{\mathcal{R}_1, \cdots, \mathcal{R}_n\}$ to $\mathcal{A}$. In addition, as $\mathcal{A}$ is a registered user, to provide a correct view for $\mathcal{A}$, $\mathcal{B}$ must also provide secret token $\tau$ and partial secret key $\eta$ for the current epoch to $\mathcal{A}$.

\noindent \underline{\textit{Query 1.}} $\mathcal{A}$ can query ciphertext and trapdoor, but $\mathcal{A}$ directly passes this phase.

\noindent \underline{\textit{Challenge.}} $\mathcal{A}$ randomly chooses two keywords $w_0, w_1$ for document $f_m$, and submits them to $\mathcal{B}$. $\mathcal{B}$ randomly chooses a bit $b \xleftarrow{\$} \{0,1\}$ and generates a challenge trapdoor $T^*_{w_b} \leftarrow \textbf{Trapdoor}(\mathbb{GP}, \mathcal{X}, y_i,I,H(\sigma_u),w_b)$ for some $i \in \{1,\cdots, n\}$, where $y_i$ is the secret key of $R_i$. $\mathcal{B}$ then returns $T^*_{w_b}$ to $\mathcal{A}$.

\noindent \underline{\textit{Query 2.}} Similar to Query 1, $\mathcal{A}$ decides to skip this phase.

\noindent \underline{\textit{Guess.}} Before outputting the result, $\mathcal{A}$ first generates a ciphertext $C_{w_0} = (C_{f,w_0}, C_{w,w_0})$ for the keyword $w_0$ through $\textbf{ConstEnc}(\mathbb{GP}, \mathcal{X}, I, \textstyle \sum_{j \in I} \mathcal{Y}_i, H(\sigma_u), f_m, w_0)$. Here we only focus on the part of $C_{w,w_0} = (C_1, C_2, C_3, C_5, C_6)$. More concretely, we have \begin{alignat*}{3}
        \beta &= H_2(\eta \mathcal{X});~~
        &&C_1 = r/\beta;
        &&C_2 = h/\beta;\\
        C_3 &= rP;
        &&C_5 = \textstyle \sum_{j \in I} r (\mathcal{Y}_i);~~
        &&C_6 = H_3(rhH(w_0)P),
    \end{alignat*} where $r, h \xleftarrow{\$} \mathbb{Z}^*_q$.
Now, $\mathcal{A}$ can follow the steps in the $\textbf{Test}$ algorithm to test whether $C_{w_0}$ and $T^*_{w_b}$ are related to the same keyword. That is, $\mathcal{A}$ computes \[\mathbb{C}_6 = H_3([(C_5 - C_1T_1) - C_3]T_2C_2)\] and check whether $\mathbb{C}_6 = C_6$. Depending on $\mathcal{B}$'s choice of $b$, two scenarios arise: \begin{itemize}
    \item If $b = 0$: we have $T^* = (T_1, T_2)$, where \[T_1 = \textstyle \sum_{j \in I \ {i}} \beta(\mathcal{Y}_i);~~~~
        T_2 = (y_i - 1)^{-1}\beta H(w_0)\] and $\beta = H_2(\eta \mathcal{X})$. Therefore, we have \begin{align*}
        \mathbb{C}_6 &= H_3([(C_5 - C_1T_1) - C_3]T_2C_2)\\
        &= H_3([r(\mathcal{Y}_i) - C_3] T_2C_2)\\
        &= H_3([r(\mathcal{Y}_i) -rP][(y_i - 1)^{-1}\beta H(w)] h/\beta)\\
        &= H_3(rhH(w_0)P) = C_6.
    \end{align*} 
    Since $\mathbb{C}_6 = C_6$, $\mathcal{A}$ outputs $b' = 0$ to indicate that $\mathcal{B}$ chooses $b = 0$ in the Challenge phase.

    \item If $b = 1$: we have $T^* = (T_1, T_2)$, where \[
        T_1 = \textstyle \sum_{j \in I \ {i}} \beta(\mathcal{Y}_i);~~~~
        T_2 = (y_i - 1)^{-1}\beta H(w_1)\] and $\beta = H_2(\eta \mathcal{X})$. Now, we have $\mathbb{C}_6 = H_3(rhH(w_1)P) \neq H_3(rhH(w_0)P) = C_6$. Therefore, $\mathcal{A}$ outputs $b' = 1$ to indicate that $\mathcal{B}$ chooses $b = 1$ in the Challenge phase.
\end{itemize}
With the above analysis, $\mathcal{A}$ can always accurately determine the value of $b$ by utilizing the ciphertext $C_{w_0}$ he/she generates. Consequently, the advantage of $\mathcal{A}$ winning the game is non-negligible.

\begin{remark}
\label{rm1}
In this section, our focus is solely on attacks launched by malicious users (\textit{i.e.,} external adversaries). We exclude attacks launched by the public cloud (\textit{i.e.,} inside adversary) from our consideration. The reason is based on the fact that the public cloud, not being authorized by the certificate authority, cannot obtain the information about secret token $\tau$ for each epoch. Consequently, we believe that as long as the public cloud does not collude with any malicious user, the public cloud is incapable of executing attacks.
\end{remark}

\section{Discussion and Solution}
\label{Discussion}
In this section, we discuss why attacks remain possible even though the security proof of Theorem~\ref{theorem} is established in~\cite{DBLP:journals/tifs/YangTZH23}. After that, we explore potential solutions to solve the security issue.

At a high level, the approach of Yang \textit{et al.} in achieving KT-IND-CKA security primarily involves preventing malicious users from having the information about the secret token $\tau$ and the partial key $\eta$,  which are necessary to generate valid ciphertexts for testing the obtained trapdoor.
However, as our cryptanalysis in Section~\ref{cryptanalysis} reveals, all registered users, including malicious ones, can access $\tau$ and $\eta$. Consequently, malicious users can effortlessly generate ciphertexts to test any obtained trapdoors, thereby compromising KT-IND-CKA security. It raises questions about the validity of Yang \textit{et al.}'s security proof. To prove Theorem~\ref{theorem}, Yang \textit{et al.}'s strategy is to embed the information of $\eta$ into the DDH problem. Specifically, in the security proof, the challenger $\mathcal{B}$ is given a DDH problem instance $(\mathbb{G}, P, \eta P, xP, zP)$, and $\mathcal{B}$'s task is to distinguish whether $x\eta P = zP$. 
$\mathcal{B}$ then initializes the system, producing $\mathbb{GP}$, and sends ``only'' the public key $(\mathcal{X}, \mathcal{Y}_1,\cdots, \mathcal{Y}_n)$ to $\mathcal{A}$. However, as $\mathcal{A}$ is a (malicious) registered user, he/she must have the information about $\tau$ and $\eta$. This means that $\mathcal{B}$ fails to correctly simulate the system for $\mathcal{A}$. Is it possible to let $\mathcal{B}$ provide such information to $\mathcal{A}$? Unfortunately,
based on the definition of the DDH problem, it is important that $\mathcal{B}$ does not have any information about $\eta$. If $\mathcal{B}$ has information about $\eta$, $\mathcal{B}$ can easily verify whether $\eta(xP) = zP$ without the need of using $\mathcal{A}$ as a subroutine.
In summary, the proof exists the following flaws: \begin{itemize}
    \item If $\mathcal{B}$ does not provide $\eta$ to $\mathcal{A}$: $\mathcal{B}$ cannot accurately simulate the system view for $\mathcal{A}$.
    
    \item If $\mathcal{B}$ provides $\eta$ to $\mathcal{A}$: It is no a correct reduction, as $\mathcal{B}$ has the knowledge of $\eta$, and can solve the DDH problem without A's involvement.
\end{itemize}

We now discuss a potential solution. To satisfy KT-IND-CKA security, DCC-SE  could be enhanced by integrating the concept of ``designated-tester PEKS''~\cite{DBLP:journals/jss/RheePSL10}. In this model, only a designated server is allowed to execute the \textbf{Test} algorithm. A straightforward approach to achieve this model is to encrypt trapdoors using the designated server's public key. This approach is akin to establishing a secret channel between each receiver and the designated server. Hence, even though any user in the system can generate ciphertexts, they cannot execute \textbf{Test} algorithm and further obtain the keyword information of the obtained trapdoors. In addition, since the designated server is not registered in the system, he/she cannot launch the attacks. Similar to Remark~\ref{rm1}, this solution cannot resist the situation in which malicious users collude with the designated tester. In this case, the designated tester can also obtain $\tau$ and $\eta$. Therefore, the designated tester can still attack KT-IND-CKA security as illustrated in Section~\ref{cryptanalysis}.

\section{Acknowledgment}
\noindent This research is partially supported by the National Science and Technology Council, Taiwan (ROC), under grant numbers NSTC 111-2221-E-004-005-, NSTC 112-2221-E-004-004-, and NSTC 112-2634-F-004-001-MBK. 

\bibliographystyle{IEEEtran}
\bibliography{bib}

\end{document}